\documentclass{jps-cp}
\usepackage{txfonts} 
\usepackage{cite}
\usepackage{amsmath,amssymb,amsfonts}
\usepackage{amsmath}
\usepackage{algorithmic}
\usepackage{graphicx}
\usepackage{textcomp}
\usepackage{xcolor}
\usepackage{mathtools}
\usepackage{multirow}
\usepackage{booktabs}
\usepackage{enumerate}
\usepackage{subfigure}
\usepackage{multirow}
\usepackage{enumitem}
\usepackage{comment}

\title{First Demonstration Experiment for Energy Trading System EDISON-X Using the XRP Ledger}
 
\author{
Yuichi \textsc{Ikeda}$^{1}$,
Yu \textsc{Ohki}$^{1}$,
Zelda \textsc{Marquardt}$^{1}$,
Yu \textsc{Kimura}$^{2}$,
Sena \textsc{Omura}$^{2}$,
Emi \textsc{Yoshikawa}$^{3}$
}
 
\inst{$^{1}$Graduate School of Advanced Integrated Studies in Human Survivability, Kyoto University, Kyoto 606-8306, Japan\\
$^{2}$CauchyE Ltd., Kyoto 602-8061, Japan\\
$^{3}$Ripple Labs Inc., California 94104, USA\\
}

\email{ikeda.yuichi.2w@kyoto-u.ac.jp}

\abst{We develop an energy trading system, EDISON-X, that uses blockchain technology to manage the buying and selling of electricity usage rights, i.e., tokens. UPX and SPX tokens purchase electricity from the utility company's distribution lines and the photovoltaic panels. On July 1, 17 students in our school dormitory participated in an experiment to confirm the operation of the EDISON-X system. Based on the results of this experiment, we discuss the energy trading system using blockchain technology for the effective usage of renewable energy. We develop topology and network science methodologies to understand the characteristics of energy trading. We test the hypothesis that market transactions become less active when ``cavities'' appear using persistent homology. The preliminary result implies that the hypothesis could be adopted. We, however, need more data samples.}

\kword{energy trading system, blockchain, token, renewable energy, demonstration experiment, hypergraph, persistent homology}

\begin{document}
\maketitle

\section{Introduction}
Blockchain can be a fundamental technology to provide solutions to various global issues: international remittance for migrants with low cost, fast and reliable digital ID for medical services for refugees, financial inclusion to provide everyone with nondiscriminatory financial services, commerce management, such as supply chain and commodity market, economic support for financing and talent matching, and trade of distributed energy, i.e., renewable energy such as solar insolation energy and wind energy \cite{Galen2019}. In addition, to use blockchain and crypto-assets to provide solutions to various global issues, the price of the crypto-assets must be stable, the amount of electrical energy required for transactions validation must be appropriate, the cost of the transaction must be low, the speed of the transaction must be high, and the occurrence of anomaly events such as money laundering and fraud must be prevented.

We focus on energy trading systems using blockchain technology, among global issues. In this energy trading, using distributed energy sources, represented by renewable energy, is essential. Blockchain technology is often utilized in energy-related use cases because it enables peer-to-peer transactions. One prime example is Brooklyn Microgrid \cite{Khalil2021}, a distributed electricity marketplace powered by blockchain technology, which allows consumers and prosumers to share electricity demand and supply information, and transact electricity with each other without an intermediary. Power ledger, an energy technology company, has also been developing a trustless, transparent and interoperable energy trading platform \cite{PowerLedger} powered by a token to align participants' incentives.

Based on these precedents, we implemented an energy trading system using blockchain technology in a dormitory of our school, where approximately 80 students reside. The current system for collecting electricity charges is somewhat complicated and burdens the students and staff in charge of accounting. To solve the problems of complexity and transparency in collecting electricity charges, we develop an energy trading system, EDISON-X, that uses blockchain technology to manage the buying (bidding) and selling (asking) of electricity usage rights, i.e., tokens, for students residing in the dormitory of our school.

In this study, we developed a blockchain-distributed energy trading system and conducted a small-scale demonstration experiment in our school dormitory. Based on the results of this experiment, we discuss the energy trading system using blockchain technology for the practical usage of renewable energy. The goal is to develop methodologies based on topology and network science to understand energy trading characteristics and predict market changes in advance.

\section{Energy Trading System EDISON-X}

\subsection{System Architecture}
We developed an energy trading system, EDISON-X, that uses blockchain technology to manage the buying (bidding) and selling (asking) of electricity usage rights (tokens). Here EDISON-X stands for energy distribution and integration systems either on ``campus'', ``small office'', ``home'', ``remote area'', ``island region'', or ``rural areas in developing countries''. EDISON-X is built on various technologies and frameworks developed by Google, Ripple Labs, and others as open source. EDISON-X uses a cloud-based database called Firestore \cite{Firestore}, built on Google cloud and has high stability. The school dormitory's electricity usage and the amount of electricity generated by the Solar photovoltaic (PV) system are monitored and controlled by Panasonic's energy server, WeLBA, which sends the data to the Firestore.

\begin{figure}[tbh]
\centering 
\includegraphics[width=0.75\textwidth]{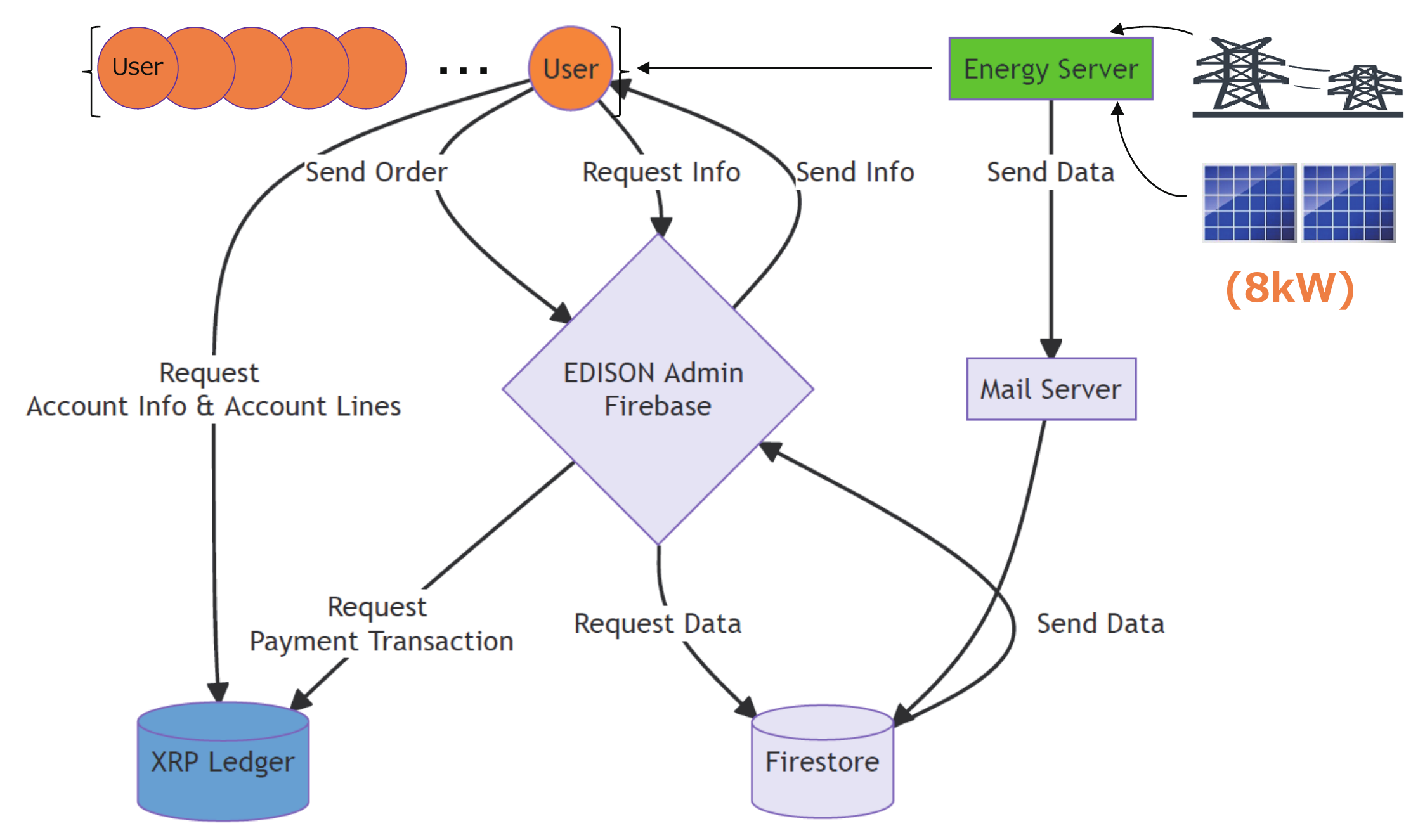}
\caption{\textbf{EDISON-X: System Configuration}, EDISON-X is built on various technologies and frameworks developed by Google, Ripple Labs, and others as open source.}
\label{f1}
\end{figure}

Figure \ref{f1} shows a system configuration of EDISON-X. We use an XRP ledger (XRPL) that consumes little electric energy for validation because it uses an algorithm instead of mining \cite{Schwartz2014, XRPL}. The XRPL is a public, permissionless blockchain that has been operating since 2012 and is one of the largest blockchain networks in the world. A key difference between XRPL and other major blockchain networks such as Bitcoin and Ethereum is that XRPL uses the Federated Byzantine Agreement (FBA) as its consensus mechanism, which makes its settlement extremely fast (3-4 seconds per settlement), low cost ($\le \$0.001$ per transaction) and scalable (1,500 transactions per second). Because of these properties, XRPL is ideal for payments and settlement use cases. XRPL also enables a wide range of other use cases, including non-fungible tokens (NFTs) and Decentralized Finance (DeFi), and became the first major blockchain to go carbon neutral in 2020. 

Any process on the Google cloud that changes the token balance is sent to the XRPL as a transaction.  Many transactions occur on EDISON-X in a short period. Therefore, the XRPL, which has a superior processing time, is adopted to solve the problem of fast transaction time. Usually, transactions are processed within seconds. These functions and the web application described below run on Firebase \cite{Firebase}, a platform on the Google cloud. Firebase has the advantage of the ease of management and high security through an authentication system developed by Google.

The buying and selling of electricity usage rights are tokens. UPX and SPX tokens purchase electricity from the utility company's distribution lines and the photovoltaic panels. UPX/SPX tokens will be issued to guarantee the right to use electricity from utility companies or photovoltaic panels. Users can buy tokens from other users who want to sell their tokens. Users can also sell their tokens to others who want to purchase them. In the case of a shortage of tokens in the entire dormitory, the system will issue additional tokens at a higher price than those issued at the beginning of the month.

\subsection{Web Application}
EDISON-X has a web application for users to access data and place orders. This application is developed in Angular. Angular is a TypeScript-based open-source web application framework led by Google. The application has three main functions. These are creating orders, retrieving data related to transactions, and checking balances. Users access Firestore through the application to check their token balance and create orders. It is also possible to see some of the information on the XRPL using the XRPL Application Programming Interface.

EDISON-X's Graphical User Interface (GUI) uses Material Design \cite{Material}. Material Design was introduced by Google in 2014 and featured simple designs inspired by materials such as paper and ink. The EDISON-X web application uses a white background and indigo as a primary color, with pink as an accent. In addition, because it is a single-page application with no page transitions on the browser, it can run as fast as a native application.

\begin{figure}[tbh]
\centering 
\includegraphics[width=0.7\textwidth]{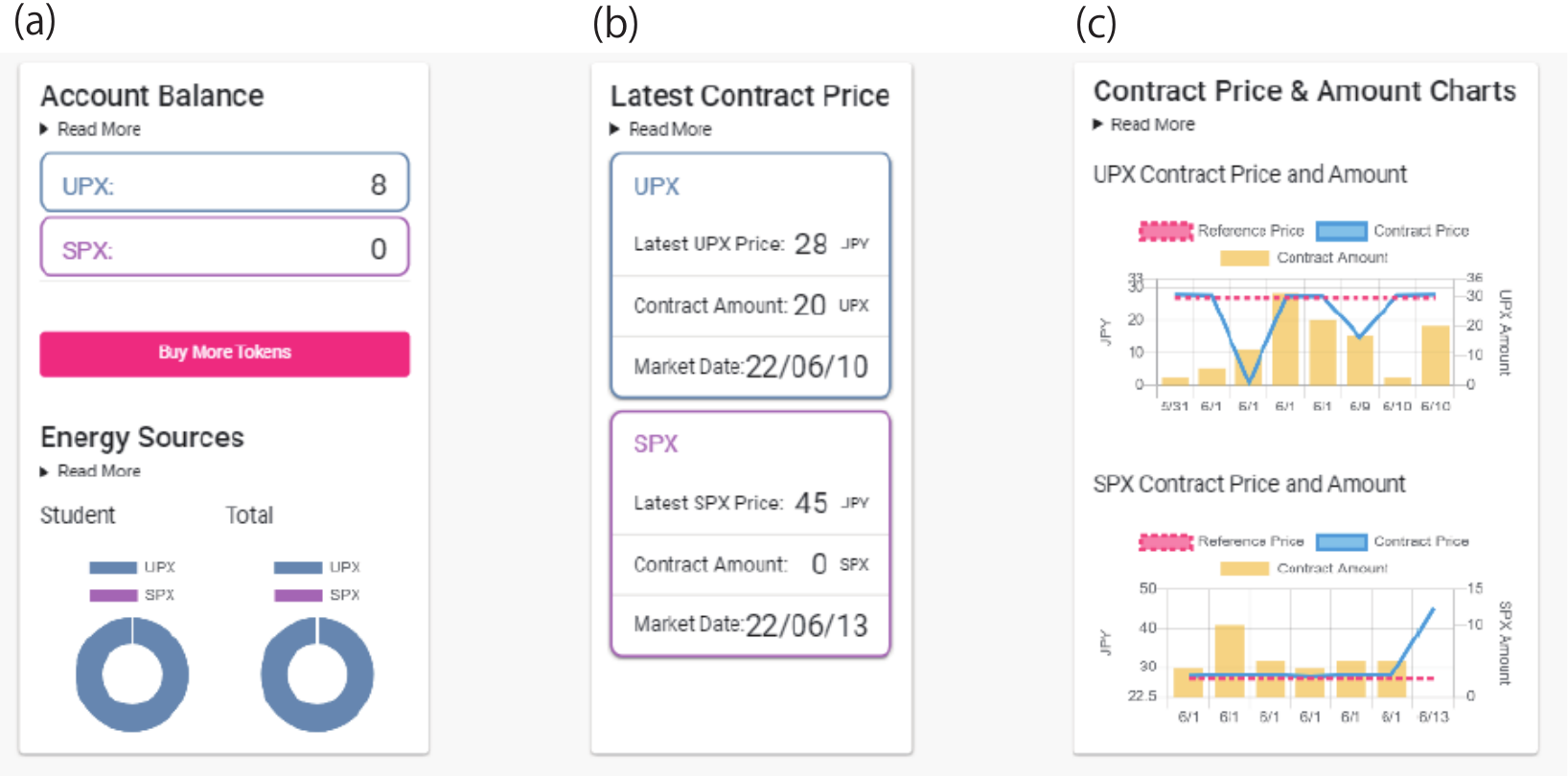}
\includegraphics[width=0.27\textwidth]{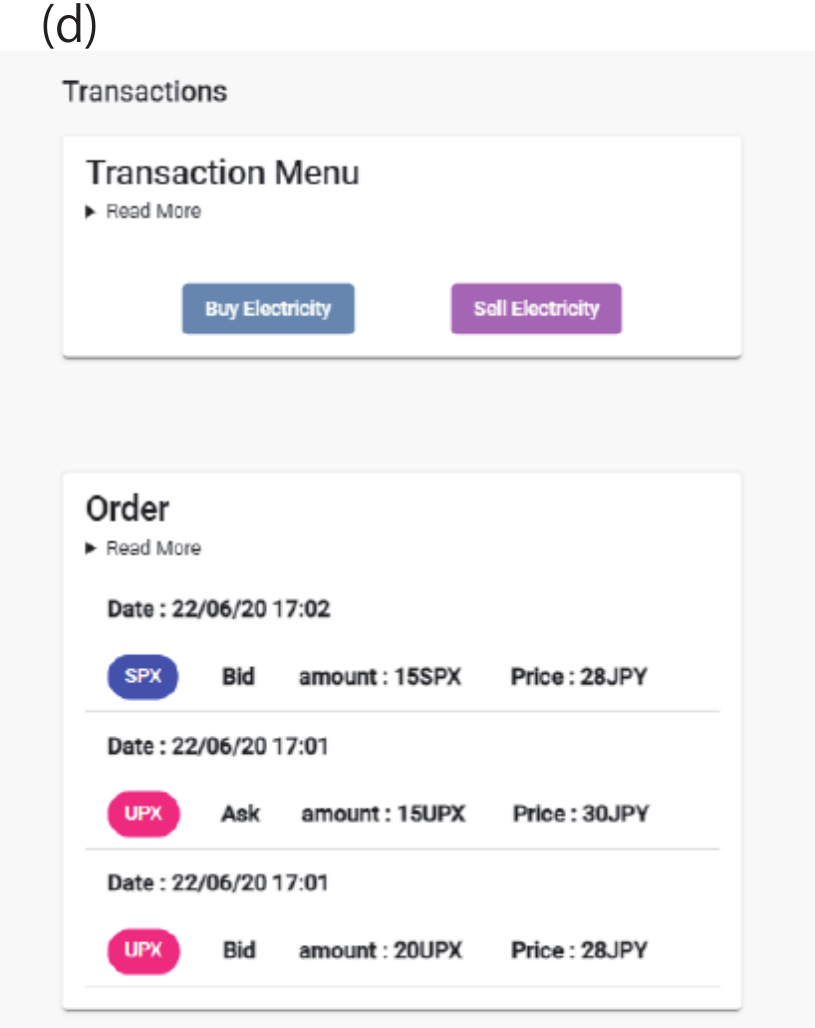}
\caption{\textbf{EDISON-X: various GUI}, Panels (a) (b) (c) show GUI Screens for viewing important EDISON-X information in one place. Panel (d) shows GUI Screens for viewing Buy/Sell Order.}
\label{f2}
\end{figure}

Figures \ref{f2} (a), \ref{f2} (b), \ref{f2} (c) show GUI Screens for viewing important EDISON information in one place. The essential information includes the ``Current account balance'', ``Last traded price'', ``This month's electricity usage'', ``Ranking of electricity usage'', ``Percentage of tokens held'', and ``Daily change of electricity usage''. Figure \ref{f2} (d) shows GUI Screens for viewing Buy/Sell Order. Participants can see the remaining available capacity (UPX, SPX, or at capacity, if any) and the latest transaction price, decide on the type, price, and quantity of tokens, and issue a buy/sell order.

\subsection{Monthly process}

\subsubsection{Beginning of the month}
The system calculates the estimated usage for this month by dividing the usage of the same month last year by the number of students. UPX/SPX tokens will be issued to guarantee the right to use electricity in the school dormitory. The price for renewable energy from solar PV will be higher, and the tokens will be sold to students who wish to use them. 

\subsubsection{During the month}
Students can buy tokens from other students who want to sell their tokens. Students can sell their tokens to other students who want to buy tokens. In the case of a shortage of tokens in the entire dormitory, the system will issue additional tokens at a higher price than those issued at the beginning of the month.

\begin{figure}[tbh]
\centering 
\includegraphics[width=0.5\textwidth]{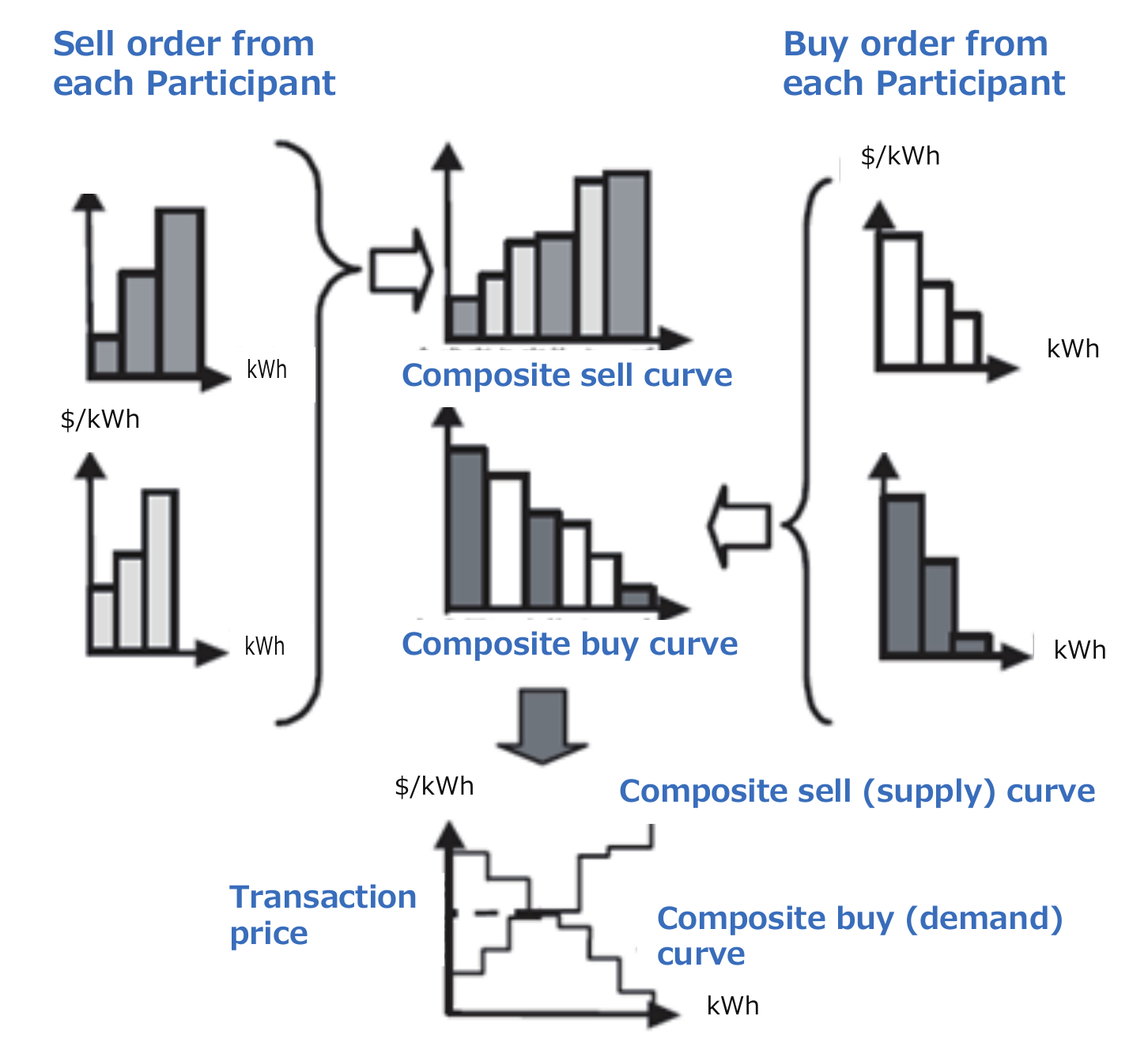}
\caption{\textbf{Token Market: Single Price Auction}, The transaction price and volume are determined at the intersection of buy order and sell order curves.}
\label{f6}
\end{figure}

Every day, the transaction price and volume are determined at the intersection of buy order and sell order curves, as shown in Fig. \ref{f6}. Buy orders higher than the transaction price and sell orders lower than the transaction price will be accepted. All orders that do not satisfy this condition are rejected.

\subsubsection{End of the month}
The system buys extra tokens at a discount from students. The system will sell additional tokens to students who need more at a higher price. The purchase and sale price will be determined so that the monthly income and expenses will be zero.

\section{Aquired Data of EDISON-X Demonstration Experiment}
Approximately 80 students reside in the dormitory of our school. The current system for collecting electricity charges is somewhat complicated and burdens the students and staff in charge of accounting. On July 1, 17 students in the dormitory of our school participated in an experiment to confirm the operation of the EDISON-X system. UPX and SPX tokens are used to purchase electricity supplied from the utility company's distribution lines and the photovoltaic panels (8kW) installed on the roof of the building of our school, respectively. We recorded the usage data of the previous month in a database. We used this data as a reference and assumed the following electricity usage for the students. The trading was carried out in a single-price auction, with the previous day's bid and ask orders being processed at midnight each day. In addition to the participants' orders, the system issued sell orders for UPX and SPX tokens when the participant's aggregated remaining token was insufficient. At the end of the month, the system buys extra tokens at a discount from students. The system will sell additional tokens at a higher price to students who need more tokens.

\begin{figure}[tbh]
\centering 
\includegraphics[width=0.65\textwidth]{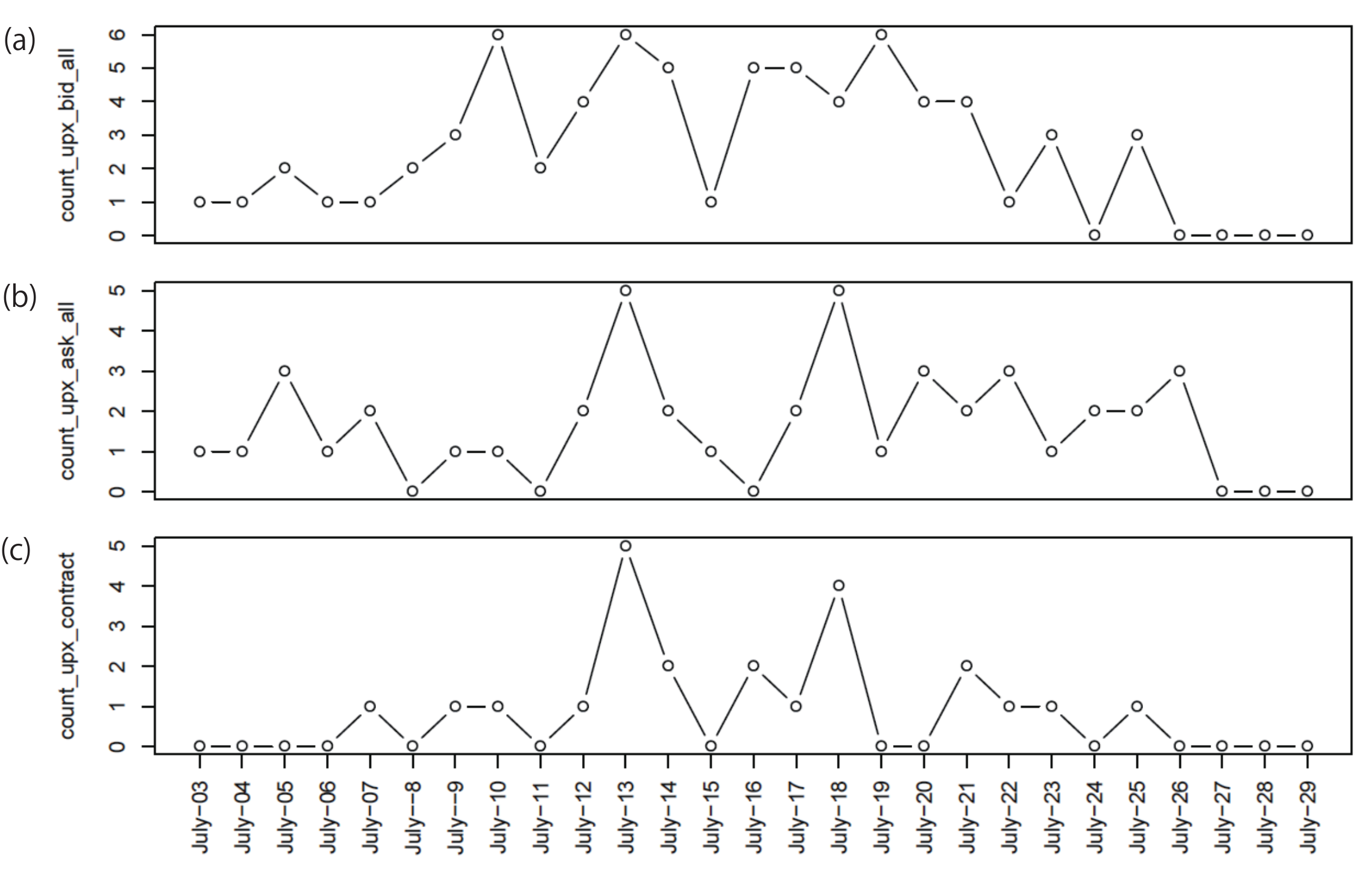}
\caption{\textbf{Bid and Ask for UPX Token}, The daily number of orders for (a) buying (bidding), (b) selling (asking), and (c) contracted buy and sell orders.}
\label{f8}
\end{figure}

Figure \ref{f8} shows the daily number of orders for buying (bidding) and selling (asking) of UPX tokens which is the right to use electricity from the utility companies. The bottom panel of Fig. \ref{f8} shows the number of contracted buy and sell orders. At the beginning and end of the month, there are fewer bids for both buying and selling orders. On average, the number of buying orders is smaller than the number of selling orders during the month.

\begin{figure}[tbh]
\centering 
\includegraphics[width=0.65\textwidth]{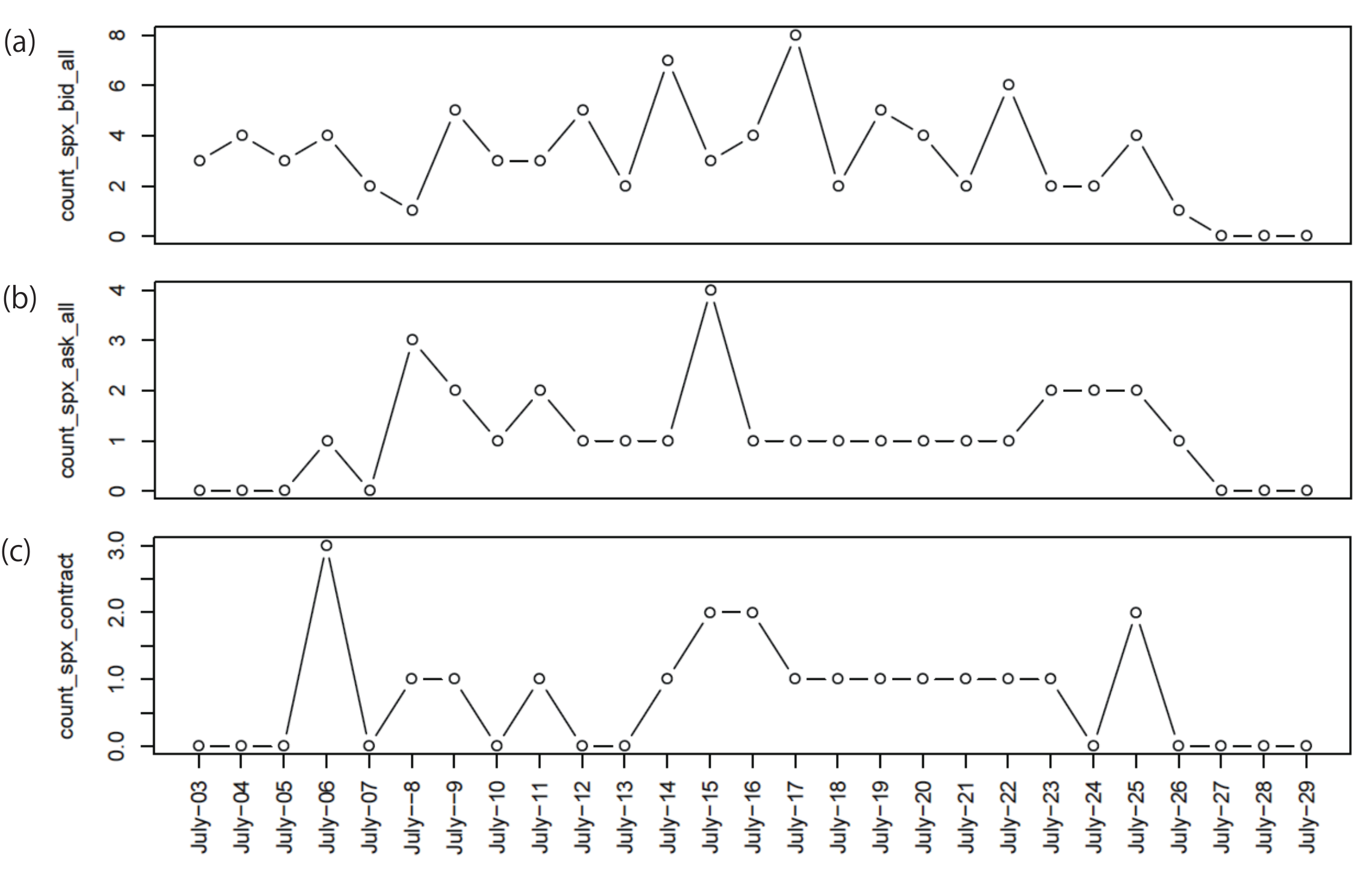}
\caption{\textbf{Bid and Ask for SPX Token}, The daily number of orders for (a) buying (bidding), (b) selling (asking), and (c) contracted buy and sell orders.}
\label{f9}
\end{figure}

Figure \ref{f9} shows the daily number of orders for buying (bidding) and selling (asking) of SPX tokens which is the right to use electricity from photovoltaic panels. The bottom panel of Fig. \ref{f9} shows the number of contracted buying and selling orders. Almost all selling (asking) orders in this experiment were placed from the EDISON-X system. The temporal variation of the number of SPX buying and selling orders shows a similar trend to that of UPX.

\begin{figure}[tbh]
\centering 
\includegraphics[width=0.4\textwidth]{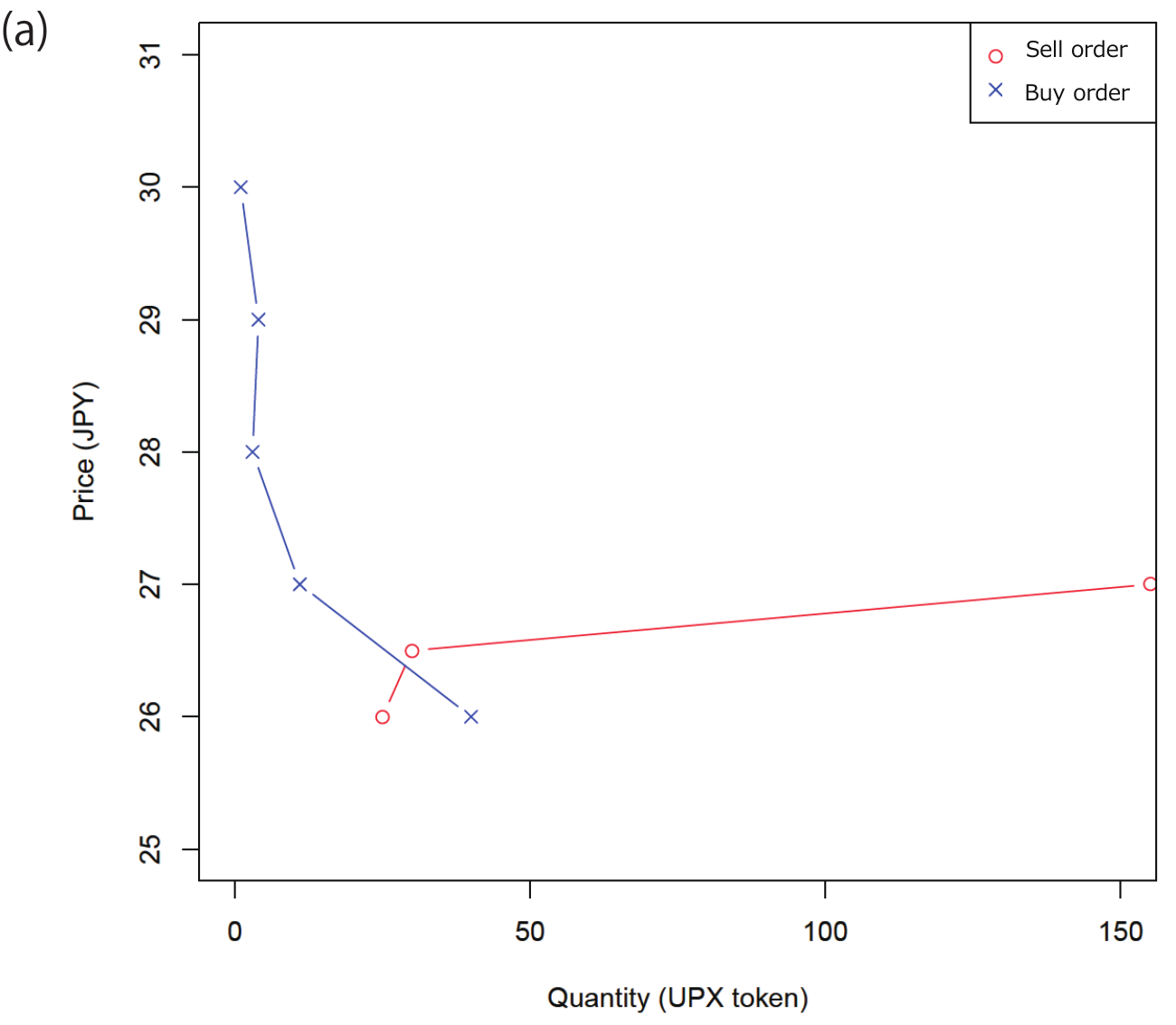}
\includegraphics[width=0.4\textwidth]{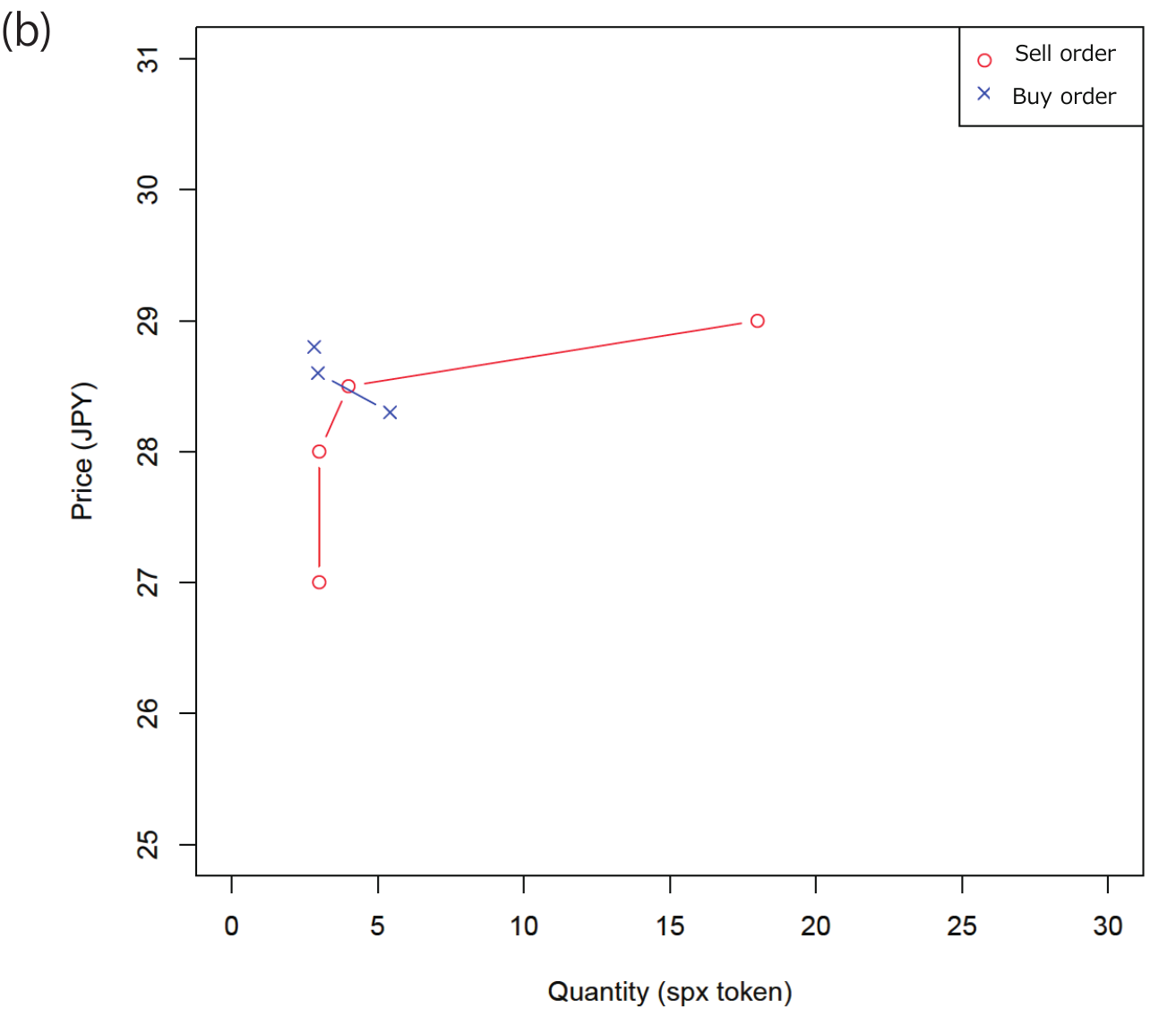}
\caption{\textbf{Typical Demand and Supply Curves}, (a) UPX Token (b) UPX Token. Transaction prices are higher on the SPX than on the UPX, and transaction volume is lower on the SPX than on the UPX.}
\label{f10}
\end{figure}

Figure \ref{f10} shows typical composite demand and supply curves. Figure \ref{f10} (a) is for the UPX Token. Figure \ref{f10} (b) is for the SPX Token. Here, the reward is given to the top SPX user for the least $\mathrm{CO}_2$ emission.  Trading price and volume are determined by the intersection of the demand and supply curves. We can see that the trading price is higher and the volume is lower on the SPX than on the UPX, reflecting that the higher trading price on the SPX is because the system sets the sell order price higher than the market price on the UPX. The high SPX transaction price reflects that the system sets the sell order price higher than the UPX market price. The low trading volume reflects the small installed capacity of solar PV.

\section{Topological Characterization of Acquired Data}

\subsection{Hypergraph of Transactions}
In an ordinary network \cite{Barabasi2016}, each edge (link) connects two vertices, but in the case of the EDISON-X energy market, we see an edge (link) must connect any number of vertices. The EDISON-X energy market is traded daily, meaning that a contract to buy or sell tokens is not concluded between two specific users but between market participants on that day. If the number of contracted orders each day equals n, the number of users involved in the contracts equals n+1. The characteristics of the EDISON-X energy market can be regarded as a higher-order network. The higher-order network is represented by a concept of topology, i.e., a simplicial complex formed by a set of simplices that is closed under the inclusion of the faces of each simplex. In network science, we often use hypergraph \cite{Bretto2013, Battiston2021, Bianconi2021}, which is an alternative representation of higher-order networks that can be used instead of simplicial complexes.

\begin{figure}[tbh]
\begin{tabular}{cc}
\begin{minipage}[t]{0.45\hsize}
\centering
\includegraphics[width=0.5\textwidth]{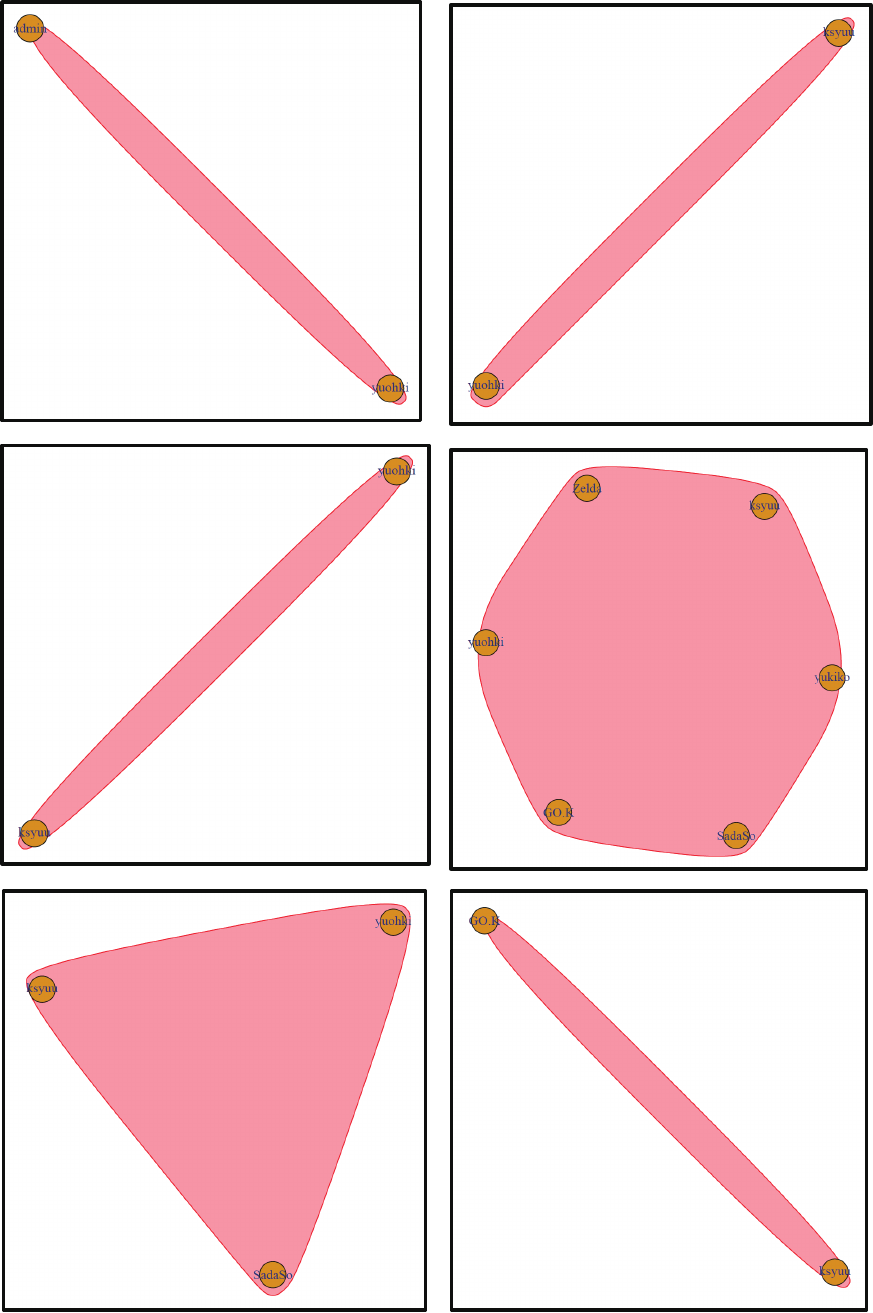}
\end{minipage} &
\begin{minipage}[t]{0.45\hsize}
\centering
\includegraphics[width=0.8\textwidth]{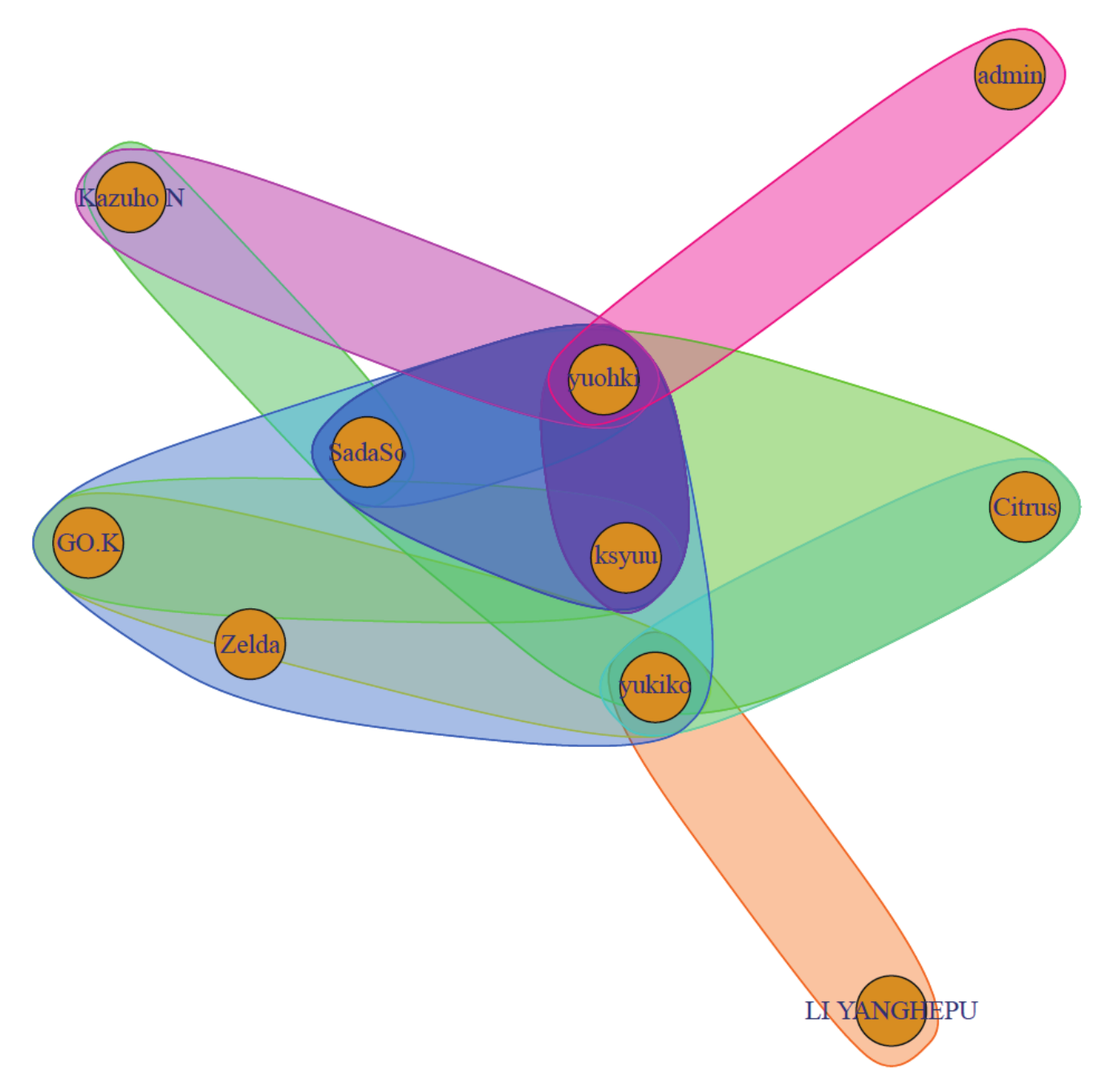}
\end{minipage}
\end{tabular}
\caption{\textbf{Hypergraph of Transactions in July for UPX Token}, The orange circles are users, and the area surrounding two or more users represents the transaction relationship. The closer the user's position is to the central part of the graph, the more important role it can be interpreted as playing in the market.}
\label{f11}
\end{figure}

In the EDISON-X energy market, transaction relationships involving more than two users can be represented by a hypergraph instead of the simplicial complexes. Figure \ref{f11} is a hypergraph showing the transaction relationship for UPX tokens in July. Figure \ref{f11} (a) shows Daily Transactions for selected days. The orange circles are users, and the area surrounding two or more users represents the transaction relationship. Transactions were concluded for four days among two users, one day among three users, and one day among six users. Figure \ref{f11} (b) shows a Hypergraph of Transactions contracted on July 07, 09-10, 12-14, 16-18, 21-23, and 25. The closer the user's position is to the central part of the graph, the more important role it can be interpreted as playing in the market.

\begin{figure}[tbh]
\begin{tabular}{cc}
\begin{minipage}[t]{0.45\hsize}
\centering 
\includegraphics[width=0.5\textwidth]{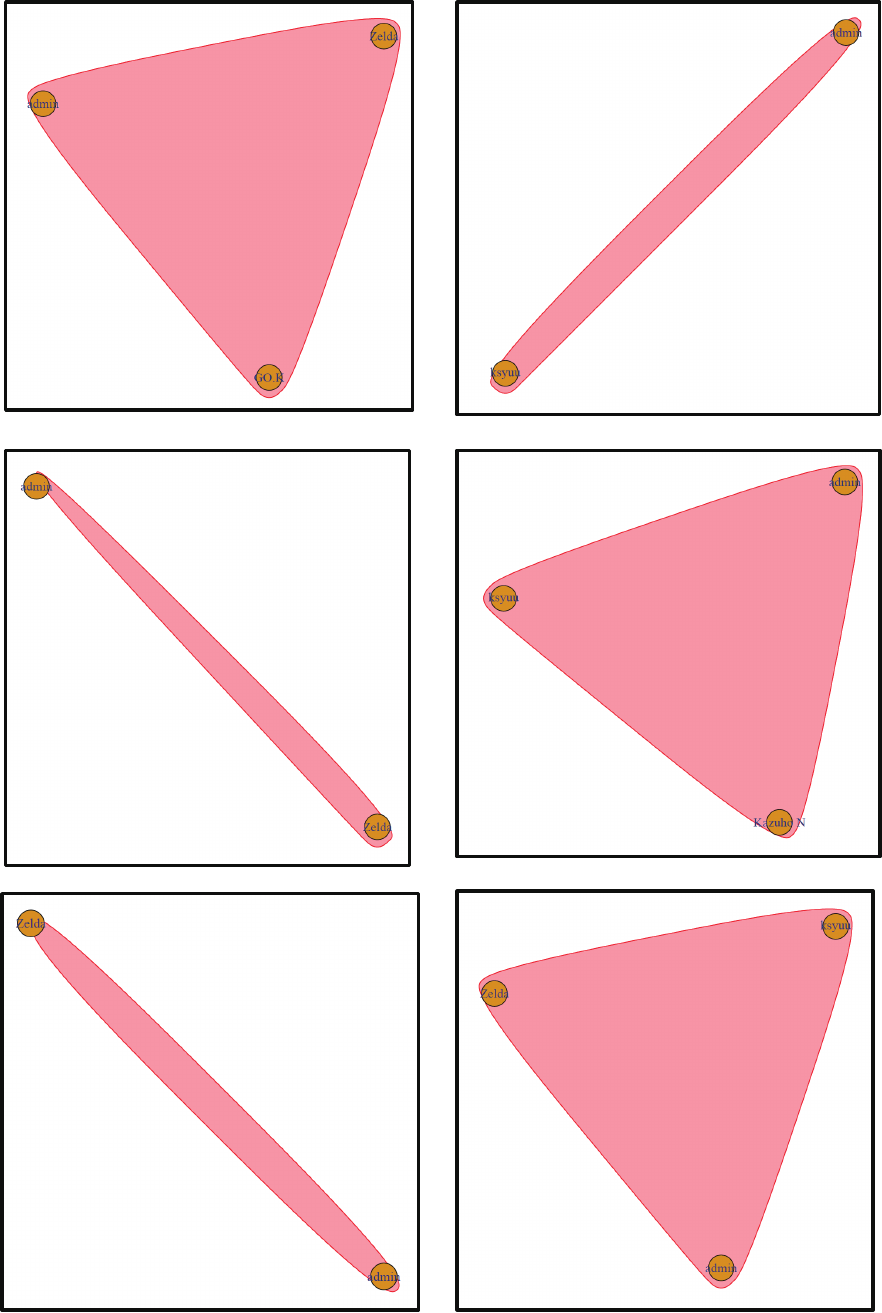}
\end{minipage} &
\begin{minipage}[t]{0.45\hsize}
\centering
\includegraphics[width=0.8\textwidth]{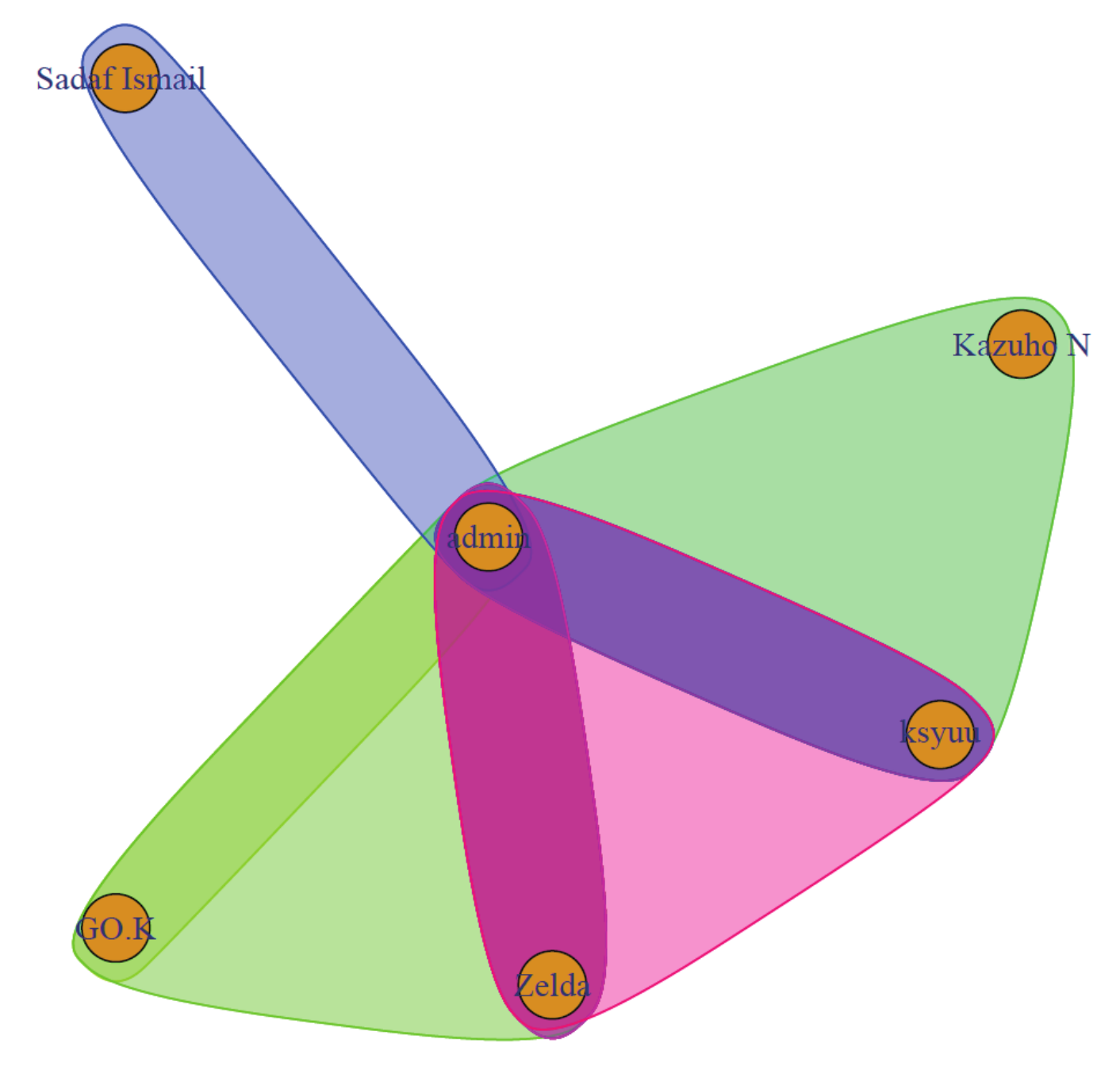}
\end{minipage}
\end{tabular}
\caption{\textbf{Hypergraph of Transactions in July for SPX Token}, Since the EDISON-X system is responsible for selling new SPX tokens, the node in the center of the graph is written ``admin'', which stands for the system administrator.}
\label{f12}
\end{figure}

Figure \ref{f12} is a hypergraph showing the transaction relationship for SPX tokens in July. The orange circles are users, and the area surrounding two or more users represents the transaction relationship. Figure \ref{f12} (a) shows Daily Transactions for selected days. Transactions were concluded for three days between two users and three days between three users. Figure \ref{f12} (b) shows a Hypergraph of Transactions contracted on July 08-09, 11, 14-23, and 25. Since the EDISON-X system is responsible for selling new SPX tokens, the node in the center of the graph is written ``admin'', which stands for the system administrator.

\subsection{Cavity Detection using Persistent Homology}
How can we characterize the ``data shape'', a feature of the distribution of many data points? It can be characterized not by ``clusters'' of data points but by the existence of ``cavities (rings, holes)''. The field of topology is useful for roughly examining such characteristics of the shape of data. In particular, cavities can be detected using homology by drawing a sphere of appropriate radius centered on each data point. Furthermore, by using persistent homology \cite{Edelsbrunner2002, Zomorodian2005}, which extends the concept of homology, it is possible to obtain not only the existence of a cavity but also detailed geometric characteristics such as its size and stability. Such a methodology is called topological data analysis (TDA) and has recently attracted much attention \cite{Edelsbrunner2010, Carlsson2009}.

We expect that market transactions will become more active when electricity consumption increases from the previous day. A disturbance in the correlation between electricity consumption and market transactions is observed as a ``cavity''. Thus, we hypothesize that market transactions become less active when ``cavities'' appear. To test this hypothesis, we quantify disturbances in the correlation between electricity consumption and market transactions by detecting one-dimensional ``cavities''.

\begin{figure}[tbh]
\centering 
\includegraphics[width=0.55\textwidth]{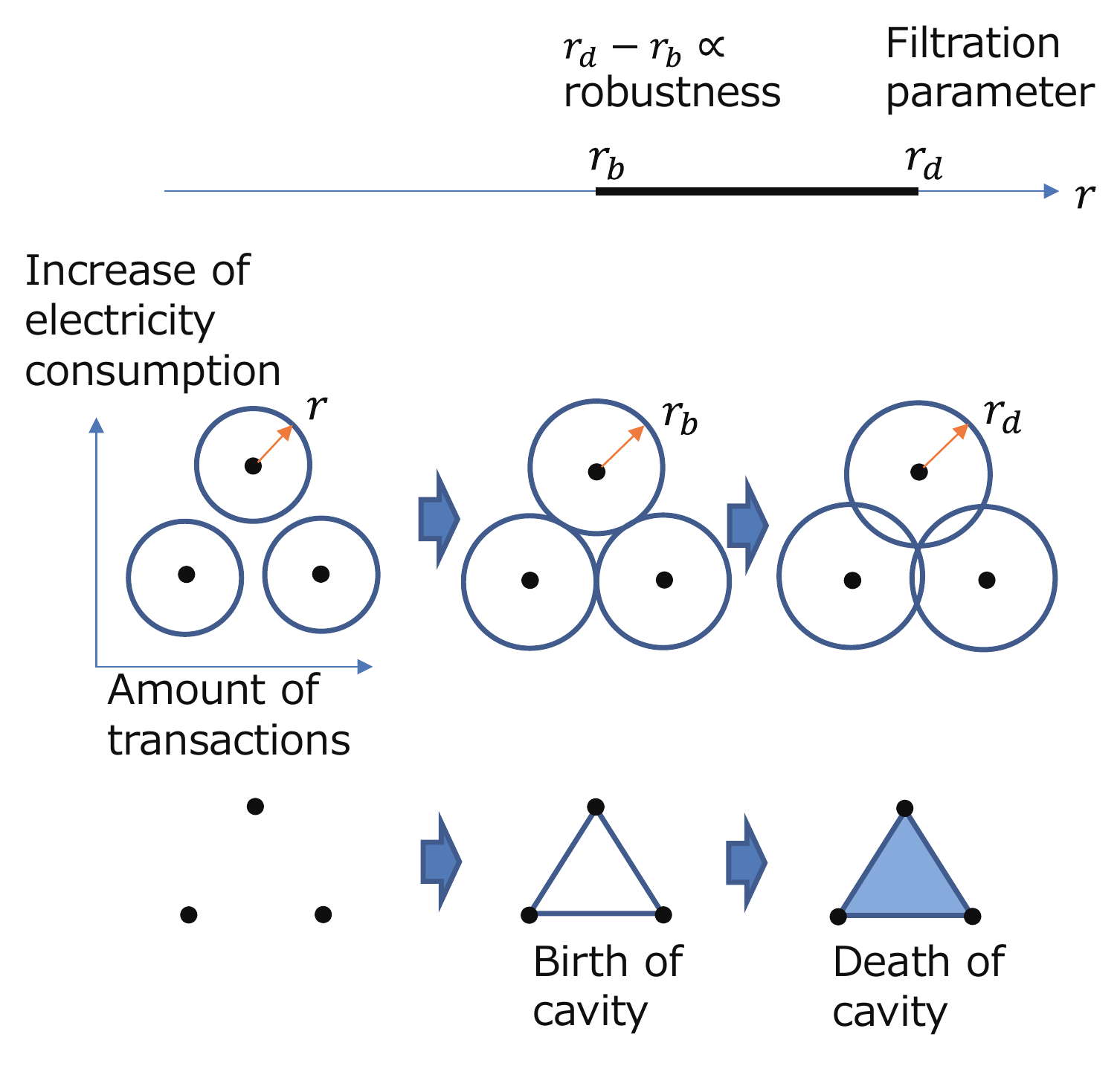}
\caption{\textbf{Persistent Homology}, The characteristics of the overall market are represented as a data point on a two-dimensional plane with the volume of transactions by each user on the x-axis and the increase in electricity usage from the previous day on the y-axis.}
\label{f13}
\end{figure}

Figure \ref{f13}  explained the concept of persistent homology. Persistent homology has been recognized as one of the essential tools of TDA. The characteristics of the overall market are represented as a data point on a two-dimensional plane with the volume of transactions by each user on the x-axis and the increase in electricity usage from the previous day on the y-axis. Draw a circle of radius $r$ from each user's data point. The circles do not overlap if the radius $r$ is small, as in the middle left panel. As the radius $r$ increases, three circles come into contact with each other at $r=r_b$, as shown in the middle panel. A cavity appears at this radius's center of the three touching circles. As the radius r is further increased, the cavity disappears at $r=r_d$ as shown in the middle right panel. The radius $r$ is called the filtration parameter, and the larger $r_d-r_b$ is, the more robust the cavity is.

\begin{figure}[tbh]
\centering 
\includegraphics[width=0.6\textwidth]{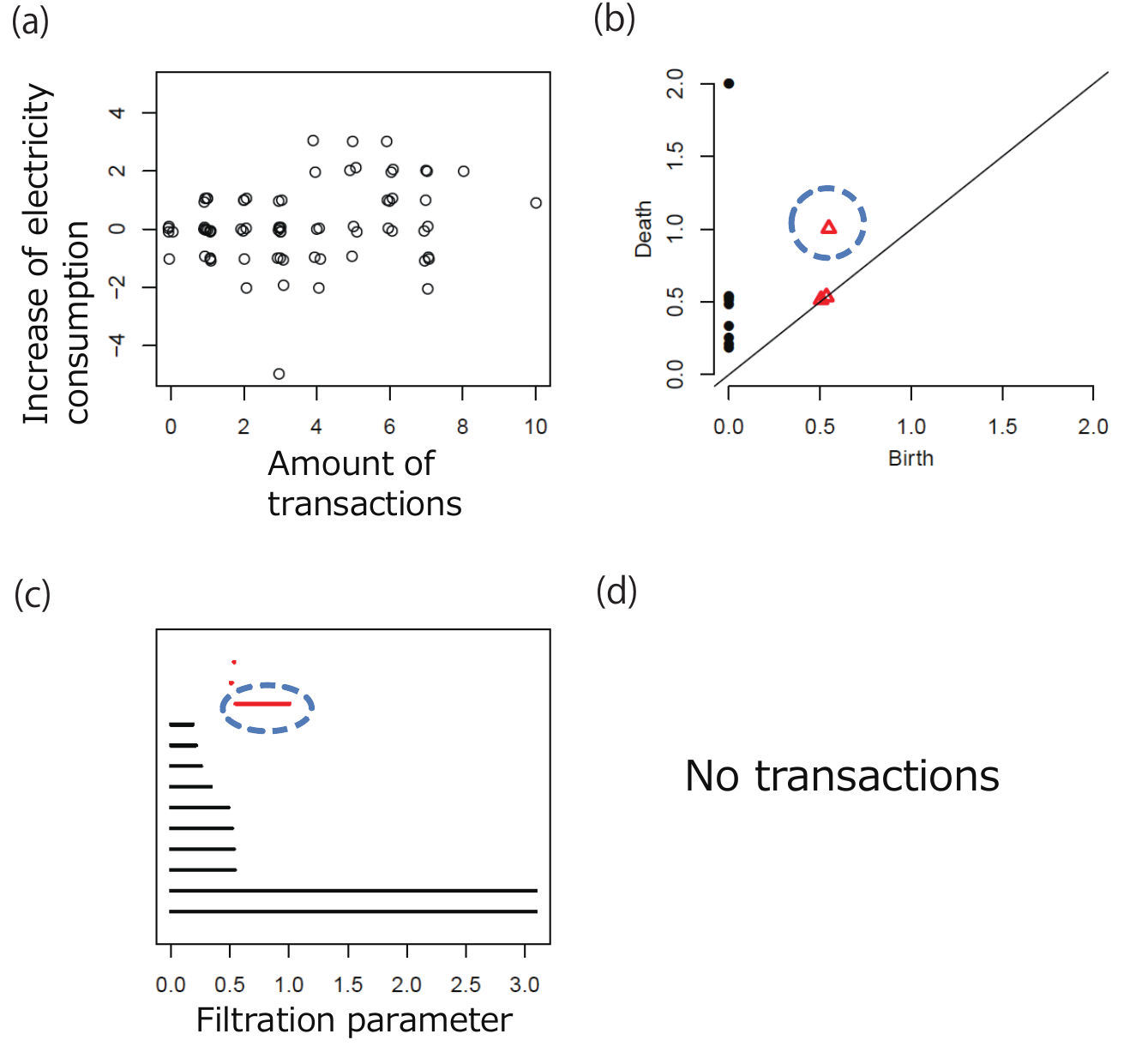}
\caption{\textbf{Market Characteristics on July 20th, 2022}, The black circles in panel (a) indicate the position of each user, and the red triangles in panel (b) indicate robust cavities. No market transactions were executed when one robust cavity was observed.}
\label{f14}
\end{figure}

\begin{figure}[tbh]
\centering 
\includegraphics[width=0.6\textwidth]{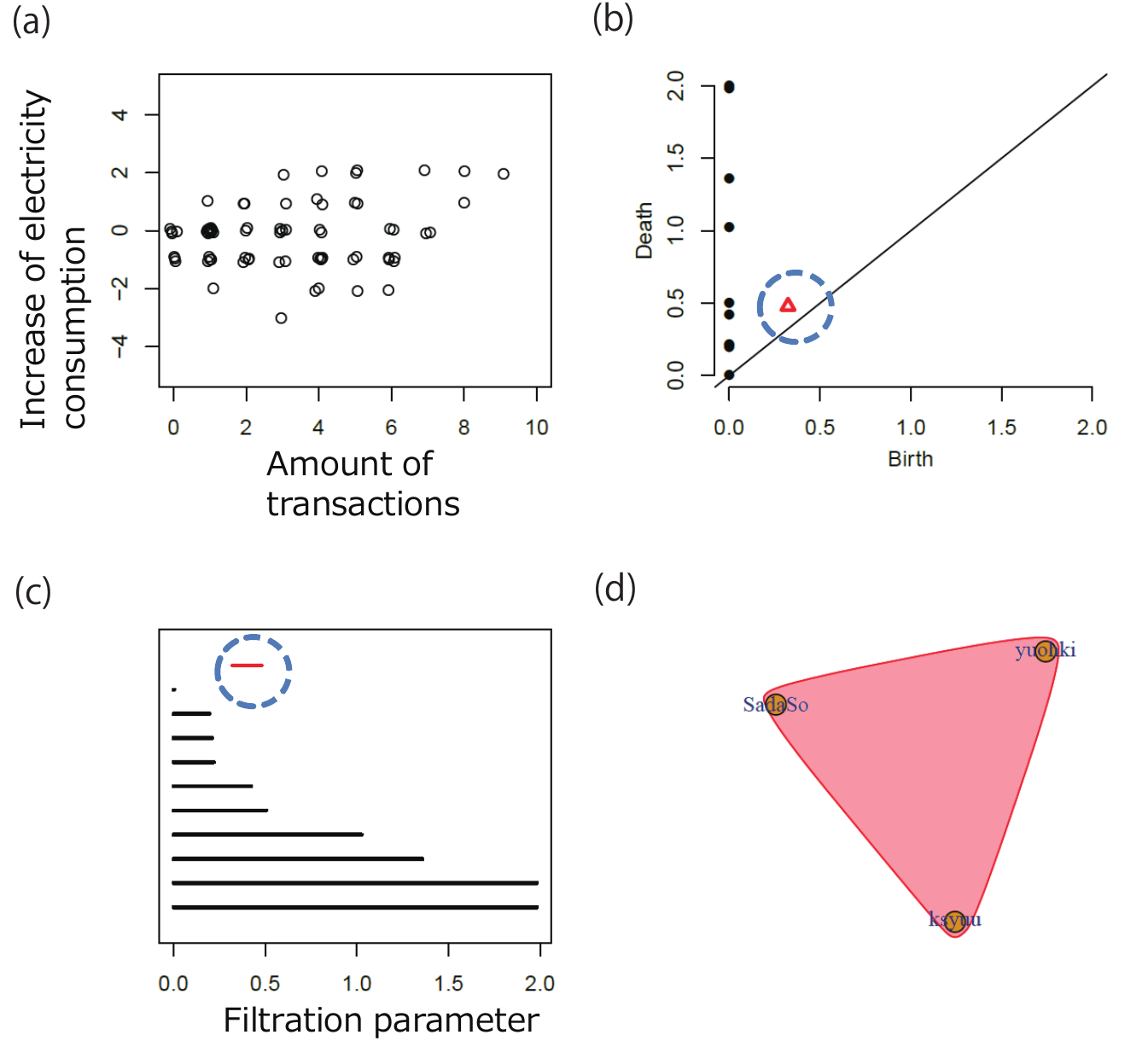}
\caption{\textbf{Market Characteristics on July 21st, 2022}, The black circles in panel (a) indicate the position of each user, and the red triangles in panel (b) indicate robust cavities. Market transactions were executed among three users, although robust cavities were observed.}
\label{f15}
\end{figure}

Figure \ref{f14} (a) shows the overall market characteristics; on July 20th, 2022, no market transactions were executed. Figure \ref{f14} (b) plots $r$ on the x-axis when the cavity appears and $r$ on the y-axis when it disappears; the further away from the 45-degree line, the more robust the cavity is. From this figure, we can see that one cavity is robust. Figure \ref{f14} (c) shows the filtration parameter from the onset to the disappearance of a cavity, indicated by the red line. The longer this line is, the more robust the cavity is. On July 20th, robust cavities were observed. Figure \ref{f15} (a) shows the overall market characteristics; on July 21st, 2022, market transactions were executed. From Fig. \ref{f15} (b) and (c),  we can see that one cavity is robust. Figure \ref{f15} (d) shows three users' transactions. On July 21st, a market transaction occurred, although robust cavities were observed.

Table \ref{t1} shows the relation of Market transactions and ``cavities''. When market transactions were executed, the number of days in which robust cavities were observed was twice (=8/4) as many as the number of days in which they were not observed. However, in the case of no market transactions, the number of days with robust cavities was 2.5 times (=5/2) the number without robust cavities. The results mean market transactions become less active when ``cavities'' appear. This result implies that the hypothesis that market transactions become less active when ``cavities'' appear could be adopted. We, however, need more data samples.

\begin{table}[tbh]
\caption{Market transactions and ``cavities'' }
\label{t1}
\begin{tabular}{lcc}
\hline
 & With transactions & No transactions \\
\hline
No cavity & 4 & 2 \\
With cavity & 8 & 5 \\
\hline
\end{tabular}
\end{table}

\section{Conclusions}
To solve the problems of complexity and transparency in collecting electricity charges, we developed the EDISON-X, which uses blockchain technology to manage the buying and selling of electricity usage rights for students residing in our school dormitory. In this study, we developed a blockchain-distributed energy trading system and conducted a small-scale demonstration experiment in our school dormitory. From July 1, 17 students in our school joined the demonstration experiment to confirm the operation of the EDISON-X system.We confirmed that the energy trading system using blockchain technology could contribute to the effective deployment of renewable energy.

We developed topology and network science methodologies to understand the characteristics of energy trading. We show that the developed methodologies help understand energy trading characteristics and predict market changes in advance. We test the hypothesis that market transactions become less active when ``cavities'' appear using persistent homology. The preliminary analysis showed market transactions become less active when ``cavities'' appear. This result implies that the hypothesis could be adopted. Currently, data acquisition and analysis are underway to get more convincing results.

\section*{Acknowledgement}
Authors thank the graduate students (alphabetical order): K. Futsuki, R. Hirata, S. Ismail, Y. Kasuya, G. Kazawa, T. Mizuguchi, Y. Morishita, T. Muraoka, K. Nomura, S. Sada, H. Sato, N. Shinohara, H. Shinto, Y. Li, and Z. Zuo, for their participants in this demonstration experiment.


\begin{thebibliography}{9}
\bibitem{Galen2019} D. J. Galen, et al., ``2019 Blockchain for Social Impact'', Stanford GSB Center for Social Innovation, September 2019.

\bibitem{Khalil2021} Khalil, S., Disfani, V., Ahmadi, D., Rollins, G., ``Impact of Blockchain Technology on Electric Power Grids - A case study in LO3 Energy'', arXiv:2106.05395, 2021. 

\bibitem{PowerLedger} Power Ledger White Paper, \texttt{https://www.powerledger.io/company/power-ledger-whitepaper}.

\bibitem{Firestore} Cloud Firestore, \texttt{https://firebase.google.com/docs/firestore, accessed 2022-11-09}.

\bibitem{Schwartz2014} D. Schwartz, N. Youngs, and A. Britto, ``The Ripple Protocol Consensus Algorithm'', Ripple Labs Inc, 2014.

\bibitem{XRPL} The XRP Ledger Foundation, \texttt{https://xrpl.org/xrp-ledger-overview.html}.

\bibitem{Firebase} Developer documentation for Firebase, \texttt{https://firebase.google.com/docs, accessed 2022-11-09}.

\bibitem{Material} Material Design, \texttt{https://m3.material.io/ Material.io, accessed 2022-11-09}.

\bibitem{Barabasi2016} A.-L. Barabasi, M. Pósfai, ``Network science'', Cambridge: Cambridge University Press. (2016). ISBN: 9781107076266 1107076269

\bibitem{Bretto2013} A. Bretto, ``Hypergraphs: Basic Concepts''. In: Hypergraph Theory. Mathematical Engineering, Springer, Heidelberg (2013). 

\bibitem{Battiston2021} F. Battiston, E. Amico, A. Barrat, et al. ``The physics of higher-order interactions in complex systems'', Nat. Phys. 17, 1093–1098 (2021). 

\bibitem{Bianconi2021} G. Bianconi. ``Higher-Order Networks, An introduction to simplicial complexes'', (Elements in Structure and Dynamics of Complex Networks), Cambridge, Cambridge University Press. 

\bibitem{Edelsbrunner2002} H. Edelsbrunner, D. Letscher, and A. Zomorodian, ``Topological persistence and simplification'', Discrete \& Computational Geometry, 28(4):511–533, Nov 2002.

\bibitem{Zomorodian2005} A. Zomorodian and G. Carlsson, ``Computing persistent homology'', Discrete \& Computational Geometry, 33(2):249–274, Feb 2005.

\bibitem{Edelsbrunner2010} H. Edelsbrunner and J. Harer, ``Computational topology: an introduction'', American Mathematical Soc., 2010.

\bibitem{Carlsson2009} G. Carlsson, ``Topology and data'', Bull. Amer. Math. Soc., 46:255–308, January 2009.


\end{thebibliography}
\end{document}